\documentclass[aps,twocolumn,nofootinbib,showpacs,superscriptaddress]{revtex4}

\usepackage{cases}
\usepackage{amsfonts}

\newcommand{\be}{\begin{equation}}
\newcommand{\ee}{\end{equation}}
\newcommand{\ba}{\begin{eqnarray}}
\newcommand{\ea}{\end{eqnarray}}
\newcommand{\nn}{\nonumber}
\newcommand{\gsim}{\raise.3ex\hbox{$>$\kern-.75em\lower1ex\hbox{$\sim$}}}
\newcommand{\lsim}{\raise.3ex\hbox{$<$\kern-.75em\lower1ex\hbox{$\sim$}}}

\def\dd{{\rm{d}}}

\def\bfxp{\vec{X}_+}
\def\bfxm{\vec{X}_-}

\def\curr{{\cal I}}

\usepackage{latexsym}
\usepackage{graphicx}
\usepackage{subfigure}
\usepackage{hyperref}

\usepackage{color}

\begin{document}

\title{Radio Broadcasts from Superconducting Strings}

\author{Yi-Fu Cai}
\email[]{ycai21@asu.edu}
\affiliation{
Physics Department, Arizona State University, Tempe, Arizona 85287, USA.
}
\author{Eray Sabancilar}
\email[]{Eray.Sabancilar@asu.edu}
\affiliation{
Physics Department, Arizona State University, Tempe, Arizona 85287, USA.
}
\author{Dani\`ele A.~Steer}
\email[]{steer@apc.univ-paris7.fr}
\affiliation{
APC\footnote{UMR 7164-CNRS, Universit\'e Denis Diderot-Paris 7,
CEA, Observatoire de Paris},
10 rue Alice Domon et L\'eonie
Duquet, F-75205 Paris Cedex 13, France.
}
\author{Tanmay Vachaspati}
\email[]{tvachasp@asu.edu}
\affiliation{
Physics Department, Arizona State University, Tempe, Arizona 85287, USA.
}


\begin{abstract}
Superconducting cosmic strings can give transient electromagnetic 
signatures that we argue are most evident at radio frequencies. 
We investigate the three different kinds of radio bursts from cusps,
kinks, and kink-kink collisions on superconducting strings. We 
find that the event rate is dominated by kink bursts in a 
range of parameters that are of observational interest, and can
be quite high (several a day at 1 Jy flux) for a canonical set 
of parameters. In the absence of events, the search for radio 
transients can place stringent constraints on superconducting 
cosmic strings.
\end{abstract}


\maketitle




\section{Introduction}
\label{intro}

Cosmic strings are one dimensional topological defects predicted in grand
unified theories (GUTs) and in superstring theory. They could be formed 
during cosmic phase transitions if the vacuum manifold associated with
the spontaneous symmetry breaking has non-trivial topology
\cite{Kibble:1976sj} (for reviews see Refs.~\cite{VilenkinBook,
Vachaspati:2006zz,Polchinski:2004ia,Copeland09,Ringeval:2010ca,Copeland11}). Since
cosmic strings are relics from the very early universe, their discovery 
could provide valuable information about the nature of fundamental 
physics.

Cosmic strings can be superconducting in a wide class of particle physics
models \cite{Witten:1984eb}, 
and can accumulate electric 
currents as they oscillate in cosmic magnetic fields, thus producing
electromagnetic effects. Oscillating superconducting strings act like
antennas of cosmic sizes that emit electromagnetic radiation
\cite{Vilenkin:1986zz,Garfinkle:1987yw,Garfinkle:1988yi} in a wide range of
frequencies from radio \cite{Vachaspati:2008su,Cai:2011bi} to gamma rays
\cite{Paczynski,Berezinsky:2001cp}. The emission is enhanced significantly 
at cusps --- where part of the string doubles on itself and momentarily
moves at the speed of light --- and at kinks --- discontinuities in the 
vector tangent to the string. 

Radiation from cusps of superconducting 
strings was suggested as the source of high redshift gamma 
rays in \cite{Paczynski}, however the burst duration turns out to be 
much smaller than that of observed gamma ray bursts. 
Strings were reconsidered as gamma ray burst engines in a
scenario \cite{Berezinsky:2001cp}
in which low frequency radiation from string cusps 
pushes the surrounding plasma, and the observed gamma ray burst 
originates as the plasma cools off  (see also 
the recent study \cite{Cheng:2010ae}). 

Recently, it was suggested \cite{Vachaspati:2008su} that
superconducting strings might best be detected in radio transient searches
since the event rate for low frequency signals is much larger than that of high
frequency signals. Furthermore, there is increasing interest in the detection
of radio transients \cite{Lorimer:2007qn,eta,lwa,lofar,ska}. More detailed
analyses were carried out in Ref.~\cite{Cai:2011bi}, where the event rate
for radio transients from cusps was obtained in terms of detector
parameters --- namely the flux, duration and frequency of the burst. In 
this paper, we re-evaluate radio transients from superconducting strings, 
taking into account signals from kinks and kink collisions. We compare 
properties and event rates of transients due to kinks with those
due to cusps. 

In addition to electromagnetic radiation, massive particles can be emitted from
superconducting strings.  Massless charge carriers are ejected from the string
when the current exceeds their mass outside the string core \cite{Witten:1984eb}.
This can occur efficiently at cusps, hence, ultra high energy neutrino
fluxes that can be observed at the future neutrino telescope JEM-EUSO, and
radio telescopes LOFAR and SKA, can be produced 
\cite{Berezinsky:2009xf}. 

The distinguishing features of bursts from superconducting cosmic string 
cusps \cite{Vachaspati:2008su,Cai:2011bi} and kinks are that string 
radio bursts are linearly polarized, and should be correlated with
gravitational wave \cite{DV} and possibly also ultra high energy 
cosmic ray bursts \cite{Berezinsky:2009xf}. Searches for correlated 
signals in these events can help distinguish their origin.
There is already an initiative for detecting the electromagnetic
counterparts of gravitational wave transients \cite{ligo-virgo}.

Cosmic strings are characterized by their tension, $\mu$, or in 
Planck units, $G \mu$, where $G$ is Newton's constant. They can 
produce a variety of observable effects, and negative results from 
cosmic string searches put constraints on $G \mu$. A bound can be
placed on the string tension from measurements of the cosmic
microwave background (CMB) anisotropies. The most recent
analysis uses the WMAP 7-year \cite{Komatsu:2010fb} and SPT 
data \cite{Keisler:2011aw}, and obtains the bound,
$G \mu \lesssim 1.7 \times 10^{-7}$ \cite{Dvorkin:2011aj}. 
Cosmic strings can also generate gravitational waves
\cite{Vachaspati:1984gt,Garfinkle:1988yi}, both in the form of bursts \cite{DV}
and a stochastic background \cite{pulsar,Olmez,Sanidas:2012ee,Dufaux12}. The
strongest constraint on $G \mu$ comes from the pulsar timing measurements that
put an upper bound on the stochastic gravitational wave background of $h^{2}\,
\Omega_{\rm{GW}} \lesssim 5.6 \times 10^{-9}$ \cite{pulsar}. Translating this
to a constraint on cosmic string tension yields $G \mu \lesssim 4 \times
10^{-9}$ \cite{pulsar}. However, since the kinetic energy of the cosmic string
loops and radiation channels other than gravitational waves have been ignored
in Ref.~\cite{pulsar}, the upper bound from pulsar timing measurements is expected
to be somewhat relaxed (see e.g., Refs.~\cite{Olmez,Sanidas:2012ee,Dufaux12}
for similar bounds). Recent measurements by WMAP \cite{Komatsu:2010fb} and SPT
\cite{Keisler:2011aw} suggest the number of relativistic degrees of freedom
at the epoch of recombination is $4$ rather than $3$ (corresponding to the
$3$ families of neutrinos). This can also be considered as a constraint on the
stochastic gravitational wave background, and yields the upper bound $G
\mu\lesssim 2 \times 10^{-7}$ \cite{Sendra:2012wh}.   

There are additional constraints on superconducting cosmic strings from the
thermal history of the universe, since such strings dump electromagnetic energy as they
decay. For redshifts $z \lesssim 10^{6}$, any form of electromagnetic energy
deposited into the universe produces spectral distortions of the CMB \cite{SZ}.
Since the double Compton and Compton scatterings that thermalize the injected
energy become inefficient at these epochs, the CMB photons cannot reach the
blackbody spectrum. The spectral measurements of the CMB by COBE-FIRAS put
upper bounds on the distortion parameters $\mu_{\rm{dist}}$ and $y_{\rm{dist}}$
\cite{cobe}, which can be translated into a constraint on the parameter space
of superconducting strings, namely, the string tension, $G\mu$, and the current
on the string, $\curr$ \cite{Sanchez89,Sanchez90,Tashiro:2012nb}. The
constraints can be even stronger if the planned CMB spectrometer project PIXIE
\cite{pixie} sees no spectral distortions \cite{Tashiro:2012nb}. Besides, the
UV photons emitted by superconducting strings can reionize neutral hydrogen,
and can effect the reionization history \cite{Tashiro:2012nv}. It was shown in
Ref.~\cite{Tashiro:2012nv} that the contribution to the ionization fraction
from strings decreases slowly whereas the reionization due to structure
formation turns on rather suddenly. This feature leads to an optical depth
different than the standard reionization scenario, hence constraints on $G\mu$
and $\curr$ can be obtained from the CMB anisotropy at large angular scales by
using the WMAP 7-year data \cite{Tashiro:2012nv}. In what follows, we choose
the string parameters $G\mu$ and $\curr$ so that they are compatible with all
the constraints mentioned above. 

This paper is organized as follows. In Sec.~\ref{burst} we calculate the
characteristics of an electromagnetic burst from a superconducting 
string cusp, kink, and kink-kink
collision. In Sec.~\ref{sec:power} we calculate the spectrum of photons and
the total electromagnetic power from cusps and kinks. In
Sec.~\ref{sec:network} we study the lifetime and number density of cosmic
string loops. In Sec.~\ref{sec:eventrate} we find the event rate in observer
variables, namely, the flux, duration, and frequency band of observation. 
We do this by calculating the Jacobian of the transformation from the 
intrinsic variables, loop length $L$, and the redshift of emission $z$, 
to the observer variables, followed by a numerical evaluation in
Sec.~\ref{sec:numerical}. We conclude in Sec.~\ref{conclusions}.

Throughout we use natural units, i.e., $\hbar = c = 1$. We also adopt
the flat CDM cosmology with $\Omega = \Omega_{\rm r} + \Omega_{\rm m}$ = 1, and
ignore the effect of the recent accelerated expansion period of the universe,
hence, set $\Lambda \approx 0$. The scale factors in the radiation and matter
eras are given respectively by $a_{\rm r} \propto t^{1/2}$ and $a_{\rm m}
\propto t^{2/3}$. The relation between the cosmological time and redshift in
the radiation and matter eras are given respectively as $t = t_{0} (1+z_{\rm
eq})^{1/2} (1+z)^{-2}$ and $t = t_{0} (1+z)^{-3/2}$. We use the values of the
cosmological parameters obtained by the WMAP satellite along with supernovae
and baryon acoustic oscillation data \cite{Komatsu:2010fb}, and take $t_{0} =
4.4 \times 10^{17}$ s, $t_{\rm eq} = 2.4\times 10^{12}$ s, $1+z_{\rm eq} =
3200$ and $1+ z_{\rm rec} = 1100$.



\section{Burst characteristics}
\label{burst}

The effective action describing a superconducting string, in which the modes
responsible for
the superconductivity are either fermionic or bosonic, is given by
\cite{VilenkinBook}
\ba
S &=& \int {\rm d}\sigma {\rm d} \tau \sqrt{-\gamma} \left\{ -\mu + \frac{1}{2} \gamma^{ab}\phi_a \phi_b  - A_\mu X^\mu_{,a} J^a \right\}
\nn
\\
&&- \frac{1}{16\pi} \int {\rm d}^4 x \sqrt{-g} F_{\mu \nu} F^{\mu \nu} \ .
\label{action}
\ea
The first term is the Nambu-Goto action, with $\gamma_{ab}$, $(a,b)=0,1$ the
induced metric on the string world-sheet and $\mu$ the string tension. The
field $\phi(\sigma,\tau)$ is a massless real scalar field living on the string
world-sheet,  the world-sheet current given by
$J^a = q \epsilon^{ab} \phi_{,b}/\sqrt{-\gamma}$ where $q$ is the charge of the current carriers, and the electromagnetic field strength is 
$F_{\mu\nu}=\partial_\mu A_\nu - \partial_\nu A_\mu$.  

Since the gravitational field of a cosmic string is characterized
by $G\mu \lesssim 1.7 \times 10^{-7}$ \cite{Dvorkin:2011aj}, it is sufficient to
consider the case of a
weak gravitational field, and here we simplify
even further to the Minkowski metric $\eta_{\mu\nu}$ since our focus is
on the classical production of bursts of electromagnetic radiation from loops
(and not gravitational wave bursts \cite{DV} nor the pair-production of photons
\cite{Steer:2010jk}).
Choosing the standard conformal gauge, the induced metric is then given by
\be
\gamma_{ab} \equiv X^\mu_{,a} X^\mu_{,b} \eta_{\mu \nu} = {\rm diag}(1,-1) \ ,
\ee
where $X^\mu (\sigma, \tau)$ is the string position, and the world-sheet
current $J^a = q (\phi',\dot{\phi})$,
where $\cdot = \partial/\partial \tau$ and $' = \partial/\partial \sigma$.
The current is conserved, $\partial_a J^a = 0$, as a consequence of the
equations of motion which read (in the Lorentz-gauge $\partial_\mu A^\mu = 0$)
\ba
\Box \phi &=& -\frac{1}{2}q\epsilon^{ab} F_{\mu \nu} X^{\mu}_{,a} X^{\nu}_{,b}
\ , 
\label{phieq}
\\
\mu \Box X^\mu &=& - F^\mu_{\; \; \sigma} X^\sigma_{,a} J^a - (\Theta^{ab}
X^\mu_{,a})_{,b} \ ,
\label{xeq}
\\
\partial_\nu \partial^\nu A^\mu &=& 
4\pi J^\mu \ , 
\label{Aeq}
\ea
where
\be
J^{\mu} = \int d^2\sigma J^a X^\mu_{,a} \delta^{(4)}(x-X(\sigma)) \ .
\label{Jmudef}
\ee
Above, $\Box = \gamma^{ab} \partial_a \partial_b = \partial^2/\partial \tau^2 -
\partial^2/\partial x^2$, and in (\ref{xeq}) $\Theta_{ab} = \phi_{,a} \phi_{,b}
- \frac{1}{2} \gamma_{ab}\phi_c \phi^c$ is the world-sheet stress energy tensor
of $\phi$.  On the right-hand-side of (\ref{xeq}) the first term is the Lorentz
force on string, which is sourced by both external electromagnetic fields as
well as those generated by the string itself through (\ref{Aeq}), while the
second term is the inertia of the current carriers. In the following, we adopt
values of parameters such that these are both negligible compared to the string
tension $\mu \sim (10^{14}\, \rm{GeV})^2$. 
In that case, (\ref{xeq}) reduces to the wave equation which is compatible
with the temporal gauge $x^0 = t = \tau$.
For a loop of invariant length $L$ in its center of mass frame, the solution of
(\ref{xeq}) is given by
\be
X^\mu(t,\sigma)= \frac{1}{2} [ X_-^\mu (\sigma_-) +
                         X_+^\mu (\sigma_+) ]\ , 
                         \label{basic}
\ee
where $\sigma_\pm = \sigma \pm t$, 
\be
{\bfxm}(\sigma_-+L) =  {\bfxm}(\sigma_-) \ , \qquad   
{\bfxp}(\sigma_++L) =  {\bfxp}(\sigma_+) \ ,
\ee
and the gauge conditions impose                     
\ba
X_-^0 &=& -\sigma_- \ ,
\nn
\\
X_+^0 &=& \sigma_+\ ,
 \nonumber \\
|{\bfxp}'| &=& 1 = |{\bfxm}'| \ .
\label{stringconstraints}
\ea

We will assume that the strings carry a current density given by
\begin{equation}
J^\mu(t,{\vec x}) = \curr \int d\sigma ~ X^\mu_{,\sigma}
                         ~ \delta^{(3)}({\vec x}-{\vec X} (t,\sigma)) \, ,
\label{jmu}
\end{equation}
where $\curr$ is the constant current on the string. The maximal value of $\curr$ 
is of order \cite{VilenkinBook}
\be
\curr_{\rm max} \; \lsim 
   \; q \sqrt{\mu} = 10^{12} \mu_{-10}^{1/2} \; {\rm GeV}\, ,
\label{currmax}
\ee
where $\mu_{-10} \equiv G \mu/(10^{-10})$.

An oscillating superconducting string loop emits electromagnetic radiation
\cite{Vilenkin:1986zz,Garfinkle:1987yw,Garfinkle:1988yi}. Just as in the case
of gravitational radiation, the power radiated in electromagnetic waves of 
frequency $\omega$ decays exponentially with $\omega$
for $\omega \gg L^{-1}$ except at cusps, kinks, and kink-kink 
collisions where bursts of beamed electromagnetic radiation
can be emitted. The case of cusps was initially studied in
\cite{Vilenkin:1986zz,Garfinkle:1987yw,Garfinkle:1988yi}, and the polarization
of the emitted beam was discussed recently in \cite{Cai:2011bi}.  Here we 
focus on kink and kink-kink collisions.

Since both $X^{\mu}$ and $J^\mu$ are periodic functions for a loop of invariant
length $L$ in its rest-frame, we work
with discrete Fourier transforms
\begin{eqnarray}
 J^{\mu}(t,\vec{x}) &=&
\sum_\omega \int \frac{d^3k}{(2\pi)^3} e^{-i(\omega t-\vec{k}\cdot\vec{x})}
J_\omega^{\mu}(\vec{k})~,
\end{eqnarray}
where $\omega=4\pi n/L $ and $n \in \mathbb{Z}^{+}$. 
On using (\ref{jmu}) and (\ref{basic}) it follows that
%
\ba
 J_\omega^\mu(\vec{k}) &=&
\frac{2 \curr}{L}
\int_0^{L/2}dt \int_0^L d\sigma
e^{i(\omega t-\vec{k}\cdot\vec{X}(t,\sigma))}X'^{\mu}(t,\sigma)
\label{Jwmu}
\nn
\\
&=&
\frac{2 \curr}{L} (I_+^\mu I_-^0 + I_+^0 I_-^\mu ) \ ,
\ea
where
\begin{eqnarray}
 I_\pm^\mu(\vec{k}) =
\int_0^{L} d\sigma_\pm e^{ik\cdot X_\pm/2} X_\pm'^{\mu} \ ,
\label{eq:Jk}
\end{eqnarray}
and $k\cdot  I_\pm^\mu(\vec{k})=0$ due to the periodicity of the loop. The
integrals (\ref{eq:Jk}) are familiar from studies of gravitational wave 
emission from oscillating string loops (see for example \cite{DV}), and 
in the $\omega L \gg 1$ limit they can
be evaluated using the standard saddle point/discontinuity approximation 
\cite{Steer:2010jk}. We briefly summarize the main results of that analysis.

Let $\vec{k} = \omega \vec{n}$ where $\vec{n}$ is a unit vector. When there is
a {\it saddle point} in the phase of (\ref{eq:Jk}), then
\be
\vec{n} = \pm \vec{X}^{\, \prime}_\pm \ ,
\label{c1}
\ee 
and expanding about this point yields (for $\omega \gg L^{-1}$)
\be
I_\pm^{\mu \; {\rm (saddle)}} \approx  
\frac{L}{(\omega L)^{1/3}} \tilde{a}\, X'^\mu_{\pm} + 
i \frac{L^{2}}{(\omega L)^{2/3}} \tilde{b} \, X''^\mu_{\pm}  + 
\ldots  \, , 
\label{saddle}
\ee
where 
we have assumed that the loops are not too wiggly so that 
$|\vec{X}_\pm''| \approx 2\pi/L$, and
\be
\tilde{a} = 
 \left( \frac{2 \pi^{1/3} }{3^{2/3}} \right)
 \frac{1}{ \Gamma(2/3)} \approx 1\, , \; \;  \tilde{b} = 
  \left( \frac{3^{2/3}}{\pi^{4/3}} \right)
  \frac{ \Gamma(2/3)}{\sqrt{3}} \approx 0.4 \ .
 \nn
 \ee
As discussed in 
\cite{Vilenkin:1986zz,Garfinkle:1987yw,Garfinkle:1988yi,BlancoPillado:2000xy}, 
slightly off the direction $\vec{n} = \pm \vec{X}_\pm'$, the integrals 
in Eq.~(\ref{eq:Jk}) acquire small imaginary components, which cause them 
to die off exponentially fast outside an angle 
\be
\theta_\omega \simeq (\omega L)^{-1/3} ~ .
\label{beamwidth}
\ee
Thus result (\ref{saddle}) is only valid in a small beam of directions about $\vec{n}$
with beam width given by $\theta_\omega$.
This beam-shape burst of radiation of frequency $\omega$ is emitted over a
duration of \cite{Paczynski}
\begin{equation}
 \delta t_\omega \simeq \frac{L^{2/3}}{\omega^{1/3}}~.
\label{beamduration1}
\end{equation}
Note that due to various effects, including the scattering of the radiation as
it travels to the observer, this is not the {\it observed} duration of the beam
\cite{Paczynski,Cai:2011bi}. Returning to (\ref{eq:Jk}), now suppose that there
is a {\it discontinuity} in $X_\pm'^{\mu}$ at some $\sigma_\pm^*$. Then in the
$\omega \gg L^{-1}$ limit, the integrals are now approximated by (see
e.g.~\cite{Steer:2010jk})
\ba
I_\pm^{\mu \rm (disc)} 
    &\approx& -\frac{2}{i\omega} \Delta X'^\mu_\pm
    \label{I-disc}
\ea
where we have neglected an overall phase, and $\Delta X'^\mu_\pm$ is 
the jump in $X'^\mu_\pm$ across the discontinuity. 


\section{Power emitted in photons}
\label{sec:power}

When both $I_{+}^\mu$ and $I_-^\mu$ have a saddle point, then this 
corresponds to a cusp since from (\ref{c1}), $\vec{n}
= \vec{X}_+' = - \vec{X}_-'$ so that $|\dot{\vec{X}}|=1$.
A saddle point in one of the integrals and a discontinuity in the 
other occurs at a kink, whereas 
a discontinuity in both corresponds to a kink-kink collision. 
In each case, the power emitted in photons per unit frequency, per unit solid
angle can be calculated through \cite{WeinbergCosmoGrav,VilenkinBook}
\be
\frac{\dd^2 P_\gamma}{\dd\omega \dd\Omega} = 
\frac{\omega^2}{2\pi} \frac{L}{4\pi} |J^\mu_\omega|^2 \, .
\label{power}
\ee


\subsection{Cusps}

The spectrum of photons from cusps can be found by substituting
Eq.~(\ref{saddle}) into Eq.~(\ref{Jwmu}), and then, using Eq.~(\ref{power}) as
\cite{Cai:2011bi,DV} 
\be
\frac{\dd^2 P_\gamma^{\rm c}}{\dd\omega \dd\Omega} \approx
    \omega^2 L \left( \frac{\curr^2}{\omega^2} \right)
\approx \curr^2 L,
  \label{Ecusp}
\ee
where we have dropped numerical factors. 
The radiation 
from a cusp is emitted in a solid angle
$\Omega \approx \theta_\omega^2 \approx (\omega L)^{-2/3}$.  
On integration, it
follows that the power emitted
is dominated by the {\it largest} frequencies, and is given by 
\be
P_\gamma^{\rm c} \approx 
\curr^2 (\omega_{\rm max} L)^{1/3} \,,
\nn
 \ee
where $\omega_{\rm max}$ can be estimated as follows \cite{Vilenkin:1986zz}.
The saddle point analysis of (\ref{eq:Jk}) shows that the dominant 
contribution is from the region around the cusp of size
$|k\cdot X_{\pm}'| < 1$, where the phase
in the integrand is not oscillating rapidly. 
This gives a time and length interval on the string world-sheet 
\be
|\Delta t|, |\Delta \sigma| 
< \frac{L}{(\omega L)^{1/3}}\ .
\label{beamduration}
\ee
Thus in one oscillation period $T=L/2$, an energy $\Delta E \sim \curr^2
(\omega_{\rm max} L)^{1/3}  L$ is radiated from a region of size given
in (\ref{beamduration}). The region itself has an energy $\approx \mu \Delta
\sigma$, and electromagnetic backreaction, which we neglect in this work, will
become important when $\Delta E \approx \mu \Delta \sigma$. This leads to 
\be
(\omega_{\rm max} L)^{2/3} \approx \frac{\mu}{\curr^2}.
\label{omegamax-cusp}
\ee
Finally, therefore, the total electromagnetic power emitted from a cusp is
\be
P_\gamma^{\rm c} = \Gamma_{\gamma}^{\rm c} \curr \sqrt{\mu} \,,
\label{pcusp}
 \ee
where $\Gamma_{\gamma}^{\rm c} \approx 10$ is found by numerical
evaluation of the power for a sample of loops \cite{Vilenkin:1986zz}.



\subsection{Kinks}

Assuming a discontinuity in $X'^\mu_+$, for a single kink event it
follows from Eqs.~(\ref{Jwmu}), (\ref{I-disc}) and (\ref{power}) that 
\ba
 \frac{\dd^2 P_\gamma^{\rm k}}{\dd\omega \dd\Omega} &\approx& 
 \frac{\curr^2 L \psi_+}{ (\omega L)^{2/3} },
\label{Ekink}
\ea
where the kink sharpness $\psi_+ = |\Delta X'^\mu_+|^2$.  Now only $I_-$ is
constrained by the saddle point condition, and given by Eq.~(\ref{saddle}), so
that for a kink the radiation is emitted in a 
``fan-shape'' set of directions of
solid angle $\Omega \approx 2\pi \theta_\omega$.  For a loop with 
$N_+ \approx
N_- \approx N$ left/right moving kinks all assumed to have a similar sharpness
$\psi$, it follows that the total power radiated is {\it independent} of
the emitted frequency, and can be calculated from Eq.~(\ref{Ekink}) as
\be\label{pkink}
P_\gamma^{\rm k} \approx \curr^2 (N \psi) \ln \left(\frac{\omega_{\rm max}}{\omega_{\rm min}}\right) \approx \curr^2 (N \psi)  \ln \left[ \sqrt{\mu} L/N \right] ,
\ee
where the upper frequency cutoff is determined by the discontinuity condition
in $I_+$ (rather than the saddle point condition in $I_-$) and is order the
inverse width of the string $\omega_{\rm max} \approx \sqrt{\mu}$. The lower
frequency cutoff is determined by the validity of the calculation leading to
(\ref{I-disc}), and can be estimated as $\omega_{\rm min} \approx (L/N)^{-1}$.
Hence the logarithmic factor can be estimated as $ \ln \left[ \sqrt{\mu} L/N
\right] \sim 100$ for a wide range of parameters.

We note that since kinks emit in a fan-shape set of directions, 
and not a narrow pencil beam, the event rate
for kink radiation will be larger than that of cusps by a factor of $(\omega
L)^{1/3} \gg 1$. However, the power emitted from a kink event is smaller than that of a cusp
for a given frequency when $N \psi \sim 1$. Hence, depending on the range of
flux and the frequencies, both kink and cusp bursts could be important for radio
transient signals. We shall discuss this issue in more detail in
Sec.~\ref{sec:analysis}.


\subsection{Kink-kink collisions}

Finally, for a single kink-kink collision, substituting (\ref{I-disc}) into
(\ref{Jwmu}), and using (\ref{power}) we find 
\be
\frac{\dd^2 P_\gamma^{\rm kk}}{\dd\omega \dd\Omega} \approx \frac{ \curr^2 L
\psi_+ \psi_- }{(\omega L)^{2}}\ ,	 
\label{Ekk}
\ee
which is radiated in all 
directions. The total power is evaluated by integrating over 
frequencies but the integral is dominated by the smallest frequency
$\omega_{\rm min} \approx N/L$.
Assuming, as above, that left 
and right moving kinks have similar magnitude sharpness $\psi$, 
and that there are approximately $N$ of each, the total power 
radiated due to kink-kink bursts is given by
\be\label{pkk}
P_\gamma^{\rm kk} \approx \frac{\curr^2 (N\psi)^2}{\omega_{\rm min} L} \approx
\curr^2  N\psi^2, 
\ee
In what follows we shall assume that $ N \psi \sim 1$, in which case, the total
power from cusps given by Eq.~(\ref{pcusp}) dominates over both the kink and
kink-kink radiation given by Eqs.~(\ref{pkink}) and (\ref{pkk}) respectively.
Hence, we shall take the total electromagnetic power emitted from cosmic string
loops to be
\be
P_{\gamma} = P_{\gamma}^{\rm c} + P_{\gamma}^{\rm k} + P_{\gamma}^{\rm kk}
\approx P_{\gamma}^{\rm c}.
\ee




\section{String network}
\label{sec:network}

\subsection{Loop lifetime}

As well as radiating electromagetic radiation, loops also emit gravitational
radiation with power
\ba
P_g = \Gamma_g G \mu,
\ea
where $\Gamma_g \approx 100$ \cite{Vachaspati:1984gt}. 
The lifetime of a string
loop can therefore be written as
\ba
\tau = \frac{\mu L}{P_g + P_{\gamma}} \equiv \frac{L}{\Gamma G\mu},
\ea
where $P_{\gamma} \approx P_\gamma^{\rm c}$, and 
\ba
\Gamma = \frac{P_g + P_\gamma^{\rm c}}{G\mu^2}\, .
\ea
A loop formed with length $L_i$ at time $t_i$ will therefore have length
\be
L(t) = L_i - \Gamma G\mu(t-t_i),
\label{length}
\ee
at time $t\geq t_i$.

%

Electromagnetic radiation becomes the dominant energy loss mechanism for loops
when $\curr >\curr_{*}$, where $\curr_{*}$ can be found from the condition
$P_{\gamma}^{c} = P_{g}$ to be
\be
 \curr_* \equiv \frac{\Gamma_g}{\Gamma_\gamma^{\rm c}} G \mu^{3/2} \approx 10^8
\mu_{-8}^{3/2}  \; {\rm GeV}. 
\ee
Thus, depending on the value of the current $\curr$, $\Gamma$ is approximately
given by
\begin{numcases}
{\Gamma \simeq}
\Gamma_g  & 
\qquad for $\curr < \curr_*$
\label{larger}
\\ 
\Gamma_g \frac{\curr}{\curr_*}  & \qquad for $\curr > \curr_*$,
\label{smallr}
\end{numcases}
where $\Gamma_{g} \approx 100$ and $\Gamma_{\gamma}^{\rm c} \approx 10$.
 



\subsection{Network evolution}


The network properties of cosmic strings have been studied in simulations
\cite{Bennett90,Allen90,Hindmarsh97,Martins06,
Ringeval07,Vanchurin06,Olum07,Shlaer10,Shlaer11}
and in analytical models
\cite{Rocha:2007ni,Polchinski07,Dubath08,Vanchurin11,Lorenz:2010sm}, and it has
been found that the network scales with the horizon. Thus, using the standard
scaling evolution for the cosmic string network, the number density of loops of
initial length between $L_i$ and $L_i + \dd L_i$ in the radiation era is given
by
\ba
\dd n({L_i},t) &\simeq& \kappa_R \frac{\dd L_i}{L_i^{5/2} t^{3/2}},
\ea
where $\kappa_R \sim 1$. Thus, on using (\ref{length}) and ignoring $t_{i}$ since $t \gg t_{i}$, 
\ba
\dd n({L},t)
&\simeq& \frac{\dd L}{(L+\Gamma G \mu t)^{5/2} t^{3/2}}, \qquad (t<t_{\rm eq})\, .
\ea
For $t > t_{\rm eq}$, the loop population contains loops that were produced in the radiation-dominated
era but survived into the matter era, as well as
loops that were produced during the matter-dominated era. 
They are expected to have a $1/L^2$ distribution, and hence
the total loop distribution is a sum of these two components, namely
\ba\label{loopmat}
\dd n(L,t) \simeq \frac{C_L }{t^2 (L+\Gamma G\mu t)^2} \dd L, \qquad (t>t_{\rm
eq}),
\ea
where
\ba
C_L \equiv  1 + \sqrt{\frac{t_{\rm eq}}{L+\Gamma G\mu t}}.
\label{cl}
\ea
Recalling that $\Gamma G \mu \leq 10^{-7}$ and today $\Gamma G \mu t_0 \lsim
10^{-2} t_{\rm eq}$, notice that the radiation era loops, and hence the 2nd term in
(\ref{cl}) will dominate for $L \ll t_{\rm eq}$.

In the following, we shall study the radio transient 
events in the matter era,
thus we are only interested in the loops that exist in the matter era. Then,
from Eq.~(\ref{loopmat}), the loop number density in terms of the redshift,
$z$, can be found as  
\ba
 \dd n(L,z) \simeq
      \frac{C_L(z) (1+z)^6 }{t_0^2 \; \left[(1+z)^{3/2} L+\Gamma G\mu t_0
\right]^2} \dd L, \; \;   (z<z_{\rm eq}),\; \;
      \label{mat_dist}
\ea
where 
 \be
C_L(z) = 1 + (1+z)^{3/4}
            \sqrt{\frac{t_{\rm eq}}{(1+z)^{3/2} L+\Gamma G\mu t_0}}.
\label{eq:CLz}
\ee


\section{Event rate}
\label{sec:eventrate}

\subsection{Burst event rate from a loop of length $L$ at redshift $z$}

Consider a loop of length $L$ at redshift $z$ with $N$ left and right-moving
kinks of typical sharpness $\psi$. 
Then, the number per unit time and per unit spatial volume of cusp, kink and
kink-kink bursts is given by 
\ba
 \dd\dot{\mathcal{N}} (L,z) \simeq
 N^p\,  \frac{(\theta_{\nu_o})^{\tilde{m}}}{L (1+z)}\,  \dd n(L,z) \, \dd V(z) ~,
 \label{ddotN}
\ea
where 
\ba
p= 0\, , && \tilde{m} = 2 \qquad {\rm for \; cusp} 
\nn
\\
p= 1\, , && \tilde{m} = 1 \qquad {\rm for \; kink} 
\nn
\\
p= 2\, , && \tilde{m} = 0 \qquad {\rm for \; kink \; kink} .
\nn
\ea
Note that the angle $\theta_\omega \sim (\omega L)^{-1/3}$ is
the {\it emitted} opening angle of the beams, so the {\it observed}
spread of the different beams is determined by 
$\theta_{\nu_o} \equiv [\nu_o (1+z) L]^{-1/3}$,
where $\nu_o$ is the observed frequency of the
burst, related to its emitted frequency $\nu_e$, by 
\be
\nu_e = \nu_o(1+z)\, .
\ee
In
(\ref{ddotN}), $\dd n(L,z)$ is the matter era loop distribution given in
(\ref{mat_dist}), while $\dd V(z)$ is the physical volume element in the matter
era given by 
\ba
\dd V (z) = 54 \pi t_{0}^{3}\, [(1+z)^{1/2} -1]^{2}\, (1+z)^{-11/2} dz \ .
\ea
Hence, the burst production rate is
\ba
\dd \dot{\mathcal{N}} (L,z)
& \simeq&
A N^p (t_0 \nu_{o}) (\nu_o L)^{m-1}\, C_L(z)
\nn
\\ & \times&
\frac{(1+z)^{m-1/2} [\sqrt{1+z}-1]^2}
     {[(1+z)^{3/2}L+\Gamma G\mu t_0]^2}\, \dd L\, \dd z ,
\label{eq:prodrate}
\ea
where we shall take $A \sim 50$ and $m=-\tilde{m}/{3}$. 


\subsection{Burst flux and duration}

For an observer, the relevant quantity is not the burst rate as a function of
loop length $L$ and redshift $z$, but rather the observed energy flux per
frequency interval, $S$, to which the instrument 
is sensitive, as well as the burst duration, $\Delta$, that can be detected. 
Thus it is necessary to transform from $(L,z)$ --- the variables occurring in
Eq.~(\ref{eq:prodrate}) --- to $(S,\Delta)$.

At a distance $r(z)$ from the loop, the energy flux per frequency interval is
obtained directly from the power radiated per unit frequency
[Eqs.~(\ref{Ecusp}), (\ref{Ekink}) and (\ref{Ekk})] for cusps, kinks, and
kink-kink bursts respectively. These expressions are averaged over a loop
oscillation period, and so one must multiply by $T_L = L/2$
and then divide by the duration of the burst $\Delta$ to get the energy flux in
the burst.
It then follows that the observed energy flux per frequency
interval, $S$ is given by
\ba
S 
& \approx & \frac{L^2  \curr^2 }{ r(z)^2 \Delta} \psi^p (\nu_o L (1+z))^{-q} \ ,
\label{Sgen}
\ea
where
\ba
p= 0\, , && q = 0 \qquad {\rm for \; cusp} 
\nn
\\
p= 1\, , && q = 2/3 \ \ \  {\rm for \; kink} 
\nn
\\
p= 2\, , && q = 2 \qquad {\rm for \; kink \; kink} \ ,
\nn
\ea
and assuming matter-dominated flat cosmology, the proper distance is
\be
r(z) = 3 t_{0} (1+z)^{-1/2}\, [(1+z)^{1/2} - 1] \ .
\label{rz}
\ee
Notice in the case of kink-kink collisions, the flux, $S$, is $L$-independent,
hence will be treated separately when calculating the transformations from
variables $(L,z)$ to $(S, \Delta)$ in Sec.~\ref{sec:analysis}.

The observed duration 
\be
 \Delta = \sqrt{\Delta{t}^2+\Delta{t}_{s}^2}~,
\label{eq:Delta}
\ee
of a burst depends on both the (observed) intrinsic duration of kink event,
$\Delta t$, 
as well as $\Delta t_s$, the contribution arising due to time delays generated
by scattering with the cosmological medium.  These are frequency dependent, and
will take a different form depending on whether we consider radio, optical, or
gamma-ray bursts.

For instance, for {\it optical} and {\it gamma ray bursts}, scattering can be
neglected, $\Delta t_s=0$. In the rest frame of the string, the intrinsic
duration of both kink and kink-kink bursts
is given by the inverse frequency of the emitted radiation, $1/\nu_e$. The
observed duration is therefore 
\be
\Delta_{\rm optical/gamma}  = \Delta t \simeq \frac{1+z}{\nu_e} 
                            = \frac{1}{\nu_o} \ ,
\label{leC}
\ee
where $\nu_o$ is the observed frequency \cite{Paczynski}.

For {\it radio bursts}, the burst duration due to scattering of radio waves
with the turbulent intergalactic
medium at given frequency, $\nu_o$, and redshift, $z$, can be modeled
as a power law
\cite{LeeJokipii1976,Kulkarnietal}
(for a review, see \cite{Rickett:1977vv})
\begin{equation}
 \Delta{t}_{s} (z) \simeq \delta{t}_1 
       \left (\frac{1+z}{1+z_1}\right)^{1-\beta}
                   \left ( \frac{\nu_o}{\nu_1} \right )^{-\beta} ~,
\label{eq:deltatS}
\end{equation}
where, the parameters are determined empirically as
\begin{equation}\label{paraexp}
\delta t_1 = 5~{\rm ms}, \  z_1 = 0.3, \ \nu_1 = 1.374 ~ {\rm GHz},
\ \beta = +4.8.
\label{emperical}
\end{equation}
Thus from (\ref{emperical}) and (\ref{leC}),
\ba
\Delta^2_{\rm radio}(z) &=&{\Delta{t}_{s}^{2} (z) + \nu_o^{-2}}
\nn
\\
&=& \frac{1}{\nu_o^2} \left[ 1+ \left\{ \left( \frac{82}{1+z}\right) \left(
\frac{\nu_1}{\nu_o}\right)\right\}^{2(\beta-1)}  \right].\, \,
\label{Delta-rad}
\ea
As $z\geq 0$, the longest radio burst has 
$\Delta_{\rm radio}^{\rm max} \equiv \Delta_{\rm radio}(z=0)$. The 
minimum burst duration will be obtained from bursts at recombination, 
when $z\simeq z_{\rm rec} \approx 1100$.



\subsection{Event rate in observer's variables}
\label{sec:analysis}

In what follows, we only focus on radio bursts, and denote $\Delta_{\rm radio}$
simply by $\Delta$.  The change of variables from $(L,z)$ to $(\Delta,S)$ can
be carried out straightforwardly.

From Eqs.~(\ref{eq:deltatS}) and (\ref{Delta-rad}) it follows that
\be
1+z = d ~(x^2 - 1)^{-1/(2\beta-2)} \ ,
\label{zofdelta}
\ee
where $d=82~\nu_1/\nu_0$ and
\be
x \equiv \Delta ~ \nu_o.
\label{xdef}
\ee
Thus $z=z(\Delta)$ so that $|\partial z/\partial S|=0$ and
\ba
 \left| \frac{\partial z}{\partial \Delta}\right| 
= \frac{\nu_o}{\beta-1} \frac{x}{(x^2 - 1)} (1+z).
\ea
Thus, the change of variables from $(L,z)$ to $(S, \Delta)$ will be given by
\ba
\dd L \dd z = \left|\frac{\partial z}{\partial \Delta} \right|
\left|\frac{\partial L}{\partial S}\right| \dd S \dd \Delta \ ,
\ea
where, unless $q=2$ (kink-kink collisions), $|{\partial L}/{\partial S}|$
can be determined from (\ref{Sgen}) since
\be
L
  =  \left[ \frac{ \nu_o^{q-1}}{\psi^p} \frac{S}{\curr^{2}}\,  r^2(z)\,
(1+z)^q\, x \right]^{1/(2-q)}, \qquad (q \neq 2).
\label{LxS}
\ee
Collecting the results together, we find
\ba\label{dLdz}
\dd L \dd z 
= \frac{\nu_o}{(2-q)(\beta-1)} \left(\frac{L}{S}\right) \frac{x}{x^2 - 1} (1+z)
\dd S \dd \Delta,
\ea
where $z=z(x)$ is given in (\ref{zofdelta}).


\subsubsection{Kinks and cusps}

The event rate for kinks and cusps as a function of flux $S$ and duration
$\Delta$ can be found by substituting (\ref{dLdz}) into (\ref{eq:prodrate})
\ba\label{Ncuspkink}
\dd \dot{\mathcal{N}} (S,\Delta)
& \simeq&
\tilde{A} \frac{N^p}{S}  [L(x,S)]^{m}   f_m(x,S)\dd S \dd \Delta,
\ea
where
\be
\tilde{A} = \frac{A t_0 \nu_{o}^{m} }{(2-q)(\beta-1)},
\ee
and
\ba
f_m(x,S) &=& \frac{ x}{x^2 - 1} C_L(z)
\nn
\\
&\times&  
\frac{(1+z)^{m+1/2}\, [\sqrt{1+z}-1]^2}
     {[(1+z)^{3/2}L(x,S)+\Gamma G\mu t_0]^2}\ ,  
\label{horror}
\ea
with $z=z(x)$ given in (\ref{zofdelta}). Now we can obtain
$L(x,S)$ that appears in Eq.~(\ref{Ncuspkink})
by using the expression Eq.~(\ref{LxS}), in which we 
substitute $r=r[z(x)]$ using (\ref{rz}) and (\ref{zofdelta}).



\subsubsection{Kink-kink bursts}

As we mentioned previously, the flux $S$ in (\ref{Sgen}), is independent 
of the loop length, $L$, for the kink-kink collisions. In this case, $z$ 
can be substituted from (\ref{zofdelta}) into (\ref{Sgen}) to give
\be
S(x) =\frac{S_0}{ P(x)},
\label{Pdef}
\ee
where
\ba
S_0 &=& \frac{\curr^2 \psi^2}{9 d\, t_0^2\, \nu_o}, 
\\
P(x) &=& x
      { (x^2 - 1)^{-\frac{1}{(2\beta-2)}} }
       { \left[\sqrt{d}(x^2 - 1)^{\frac{1}{(4\beta-4)}} - 1 \right]^2}.\,\,\,\,\,
       \label{Skk}
\ea
%
The transformation from $z$ to $S$ can be done in two steps. First, we
transform from $z$ to $x$ by using Eq.~(\ref{zofdelta}),
\be\label{dzdx}
\dd z = \left| \frac{\partial z}{\partial x}\right| \dd x = \frac{(1+z)}{(\beta-1) (x^{2} - 1)}\, x \, \dd x \ ,
\ee
 and then, from $x$ to $S$ by using (\ref{Pdef})
\be\label{dxds}
\dd x =  \left| \frac{\partial x}{\partial S}\right| \, \dd S  
= \frac{S_{0}}{|\dd P/\dd x|} \, \frac{\dd S}{S^{2}} \ .
\ee
The event rate for kink-kink collisions can be found from
Eq.~(\ref{eq:prodrate}) by using Eqs.~(\ref{dzdx}) and (\ref{dxds}) as
\ba\label{NkkL}
\dd \dot{\mathcal{N}} (S)
& \simeq&
\frac{A N^2\, t_0 \, S_{0}}{(\beta -1)} \frac{x}{(x^{2} -1)|\dd P/\dd x|} \frac{\dd S}{S^{2}}
\nn
\\
& \times& \sqrt{1+z}\, [\sqrt{1+z}-1]^2 
\nn
\\ & \times&\int \frac{C_{L}(z) \, \dd L}{ L \, [(1+z)^{3/2}L+\Gamma G\mu t_0]^2} \ .
\ea 
The integral over $L$ can be estimated as 
$\sim (1+z)^{3/4}\,\, t_{\rm eq}^{1/2}/(\Gamma G \mu t_{0})^{5/2}$. 
Hence, the
event rate for kink-kink collisions is 
\ba\label{Nkk}
\dd \dot{\mathcal{N}} (S)
& \simeq&
\frac{A N^2\, t_{\rm eq}^{1/2} \, S_{0}}{(\Gamma G \mu)^{5/2}\, t_{0}^{3/2}
(\beta -1)} \frac{x}{(x^{2} -1)|\dd P/\dd x|} 
\nn
\\
& \times& (1+z)^{5/4}\, [\sqrt{1+z}-1]^2 \, \frac{\dd S}{S^{2}} \ ,
\ea 
where $z$ can be found from (\ref{zofdelta}) in terms of $x$, and $x$ can be
solved in terms of $S$ from the polynomial given by Eq.~(\ref{Pdef}). Therefore,
the event rate can be expressed as a function of flux, $S$. However, these
steps cannot be done analytically and numerical solutions are needed. 

In the next section, we shall find the event rate for cusps, kinks and kink-kink collisions numerically.   


\section{Numerical Estimates}
\label{sec:numerical}

After having obtained the expressions for event rates of radio bursts emitted
from string cusps and kinks given by Eq.~(\ref{Ncuspkink}), and from kink-kink
collisions given by Eq.~(\ref{Nkk}), we numerically evaluate and integrate
these differential forms, and find the event rates as functions of the
observable and theoretical parameters.

\subsection{Observable parameters}

We consider the main observable parameter as the flux $S$. For our numerical
estimates we assume the string parameters
\begin{equation}
\curr = 10^5 ~{\rm GeV} \ , \ \
G\mu = 10^{-10}.
\end{equation}
Our choice of $G\mu \sim 10^{-10}$ corresponds to a symmetry breaking energy
scale of $\sqrt{\mu}\sim 10^{14}~{\rm GeV}$ at which grand unification may
occur.


We shall assume a range of observable parameters, $S$, $\nu_o$ and $\Delta$
motivated by experiments. For example, the Parkes survey can probe the
following ranges of the parameters \cite{Lorimer:2007qn},
\begin{eqnarray}
\nu_o &\in& (1.230,1.518) ~{\rm GHz}\ , \nonumber \\
\Delta &\in& (10^{-3},1) ~{\rm s} \ , \nonumber \\
S &\in& (10^{-5},10^{5}) ~{\rm Jy} \ ,
\label{eq:ranges}
\end{eqnarray}
where Jy is the unit of flux used in radio astronomy, which can be converted
into the cgs units as
\begin{equation}
1~{\rm Jy} = 10^{-23} \frac{\rm ergs}{\rm cm^2-s-Hz}.
\end{equation}

\begin{figure}
\includegraphics[scale=0.35]{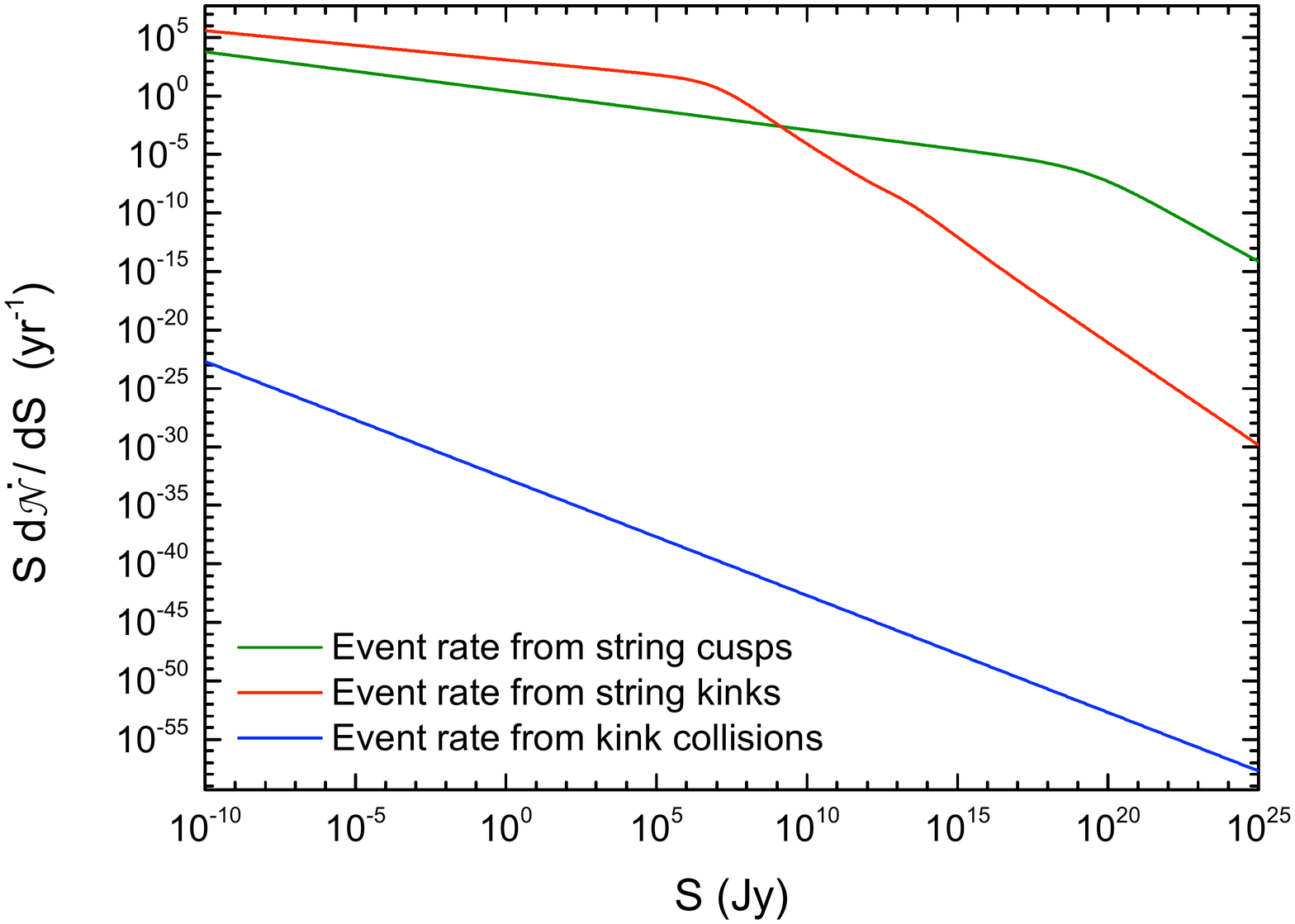}
\caption{The event rate of radio bursts from superconducting string loops with
fixed observed frequency, $\nu_o = 1.23 {\rm GHz}$ as functions of the flux
$S$. 
The green curve corresponds to the case of cusps (the top line in large
$S$ regime); the red curve corresponds to the case of kinks (the middle curve
in large $S$ regime); and the blue curve corresponds to the case of kink-kink
collisions (the bottom curve in large $S$ regime), respectively. 
We have assumed a single kink on a loop ($N=1$)
for these plots. For a different value of $N$, the plots should be
rescaled by $N$ for kinks and $N^2$ for kink-kink collisions.
}
\label{Fig:Event_S}
\end{figure}

With the above parameters, we numerically calculate the event rates of radio
bursts produced by cusps, kinks and kink-kink collisions as functions of the
flux, $S$. By integrating out the observed duration, $\Delta$, for cusps and
kinks in Eq.~(\ref{Ncuspkink}), and integrating over the loop length, $L$ in
Eq.~(\ref{NkkL}), for kink-kink collisions, the event rate per flux are
obtained numerically as a function of the observed flux, $S$, as shown in
Fig.~\ref{Fig:Event_S}.
We use log-log scale to show the wide range of scales for the flux $S$ in
Fig.~\ref{Fig:Event_S}, where it can be seen that the event rate per flux has a
power law behavior. For large $S$ values, the event rate from the string cusps
is dominant, then is the events from kinks. The contribution of the kink-kink
collisions always remains negligible compared to signal from cusps and kinks.
As $S$ decreases, the slopes of the curves for cusps and kinks change at
certain points, and correspondingly the event rate of kinks catches up and
becomes dominant at relatively smaller values of $S$. The blue (the bottom)
curve corresponding to kink-kink collisions is the most steep, and thus, one
may expect the contribution of kink-kink collisions would be the most important
when $S$ is extremely small. However, the parameter space which is detectable 
by experiments only corresponds to the range of the fluxes where most of the
events are due to kinks.

From Fig. \ref{Fig:Event_S}, we find the asymptotic power law fits for the
event rate of radio bursts emitted from cusps, kinks and kink-kink collisions
as follows, 
\begin{numcases}
{S\, \frac{d\dot{\mathcal{N}}}{dS} \simeq}
2.8\,  S_{\rm Jy}^{-1/3}\, \, {\rm yr}^{-1}~, 
& \hskip -0.2in {\rm (cusps)}\,  
\\
1.3\times 10^{3}\, N \, S_{\rm Jy}^{-1/4}\, \, {\rm yr}^{-1}~, 
&  \hskip -0.2in {\rm (kinks)}\, 
\\
2.0 \times 10^{-33}\, N^2 \, S_{\rm Jy}^{-1}\, \, {\rm yr}^{-1}~,
&  \hskip -0.2in {\rm (k-k)}
\end{numcases}
where $S_{\rm Jy} \equiv S/(\rm{1\, Jy})$ and $N$ is the typical
number of kinks per loop (conservatively taken to be one in the
plot). 
Hence, an experiment that integrates
events over the ranges of $\Delta$ in Eq.~(\ref{eq:ranges}), and is sensitive
to milli Jansky fluxes, will observe about one hundred radio bursts per day
from kinks, about one event per day from cusps, and kink-kink collisions cannot
produce observable events if there are superconducting cosmic strings with the
chosen parameters. If such events are not seen in a search for cosmological
radio transients, we will be able to place stringent constraints on 
superconducting cosmic string parameters. If we consider radio bursts 
emitted by kinks on superconducting
strings with observable frequency $1.23 {\rm GHz}$ and flux greater than
$300~{\rm mJy}$, the event rate is about 0.75 per hour, which is a factor of
$30$ larger than the upper bound given by the Parkes survey, $0.025$ per hour.
This result implies that current radio experiments might already rule out an
interesting part of parameter space given by the current on the string, $\curr$
and the string tension, $G\mu$.


\subsection{Theoretical parameters}

In this section, we numerically study the event rates as functions of
theoretical parameters, i.e., the current, $\curr$, and the string tension, $G
\mu$.

\begin{figure}[t]
\includegraphics[scale=0.35]{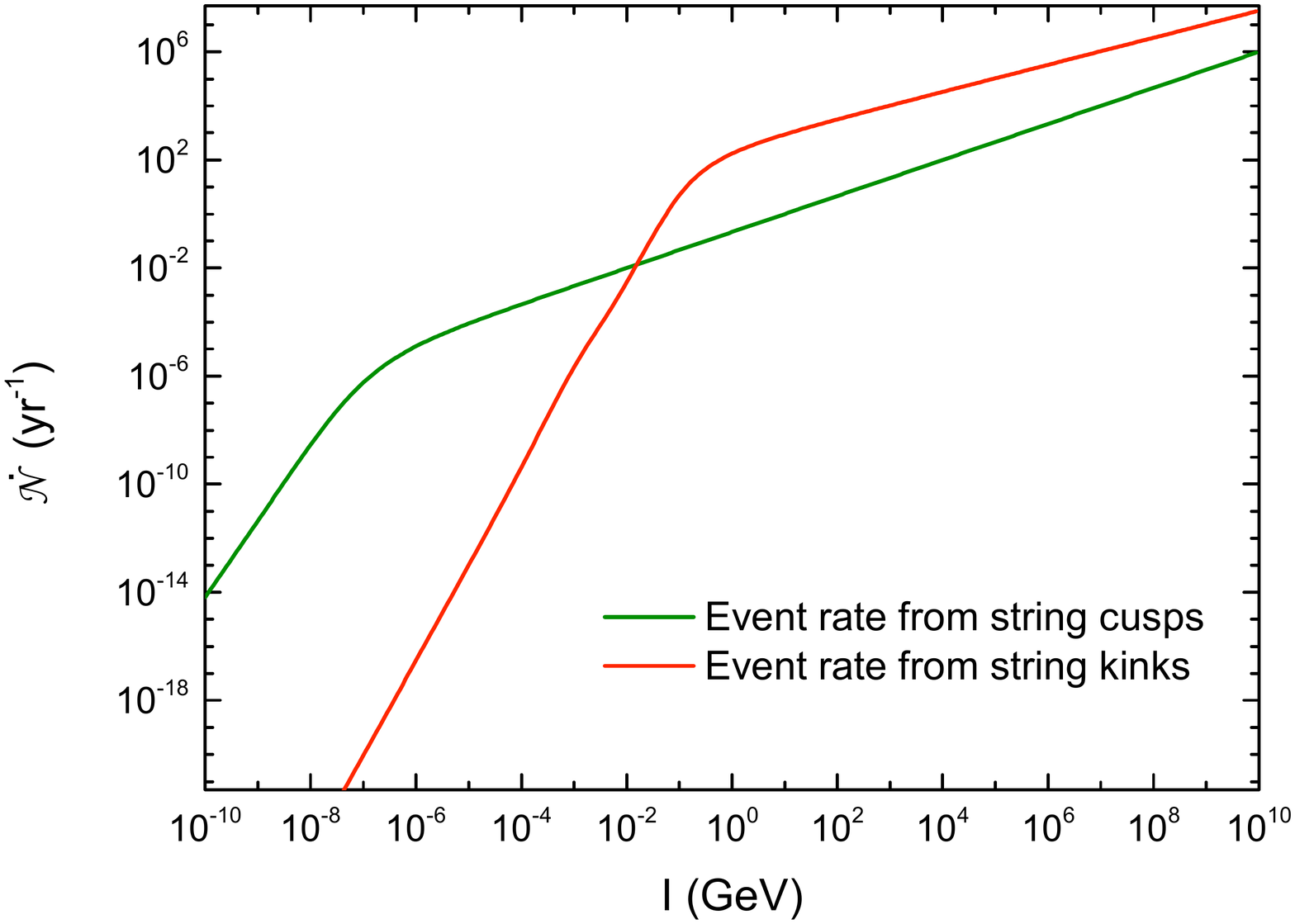}
\caption{The event rates of radio bursts from cusps 
and kinks (for one kink/loop) on superconducting string loops at fixed observed 
frequency, $\nu_o = 1.23 {\rm GHz}$, as functions of the current
$\curr$. The string tension is taken to be $G\mu=10^{-10}$.
The green curve corresponds to 
the case of cusps and gives a smaller event rate at high $\curr$; 
the red curve corresponds to 
the case of kinks and gives a smaller event rate at lower values
of $\curr$. 
The event rates for kinks
should be rescaled by $N$ if there are $N$ kinks
per loop.
}
\label{Fig:Event_I}
\end{figure}
\begin{figure}[t]
\includegraphics[scale=0.35]{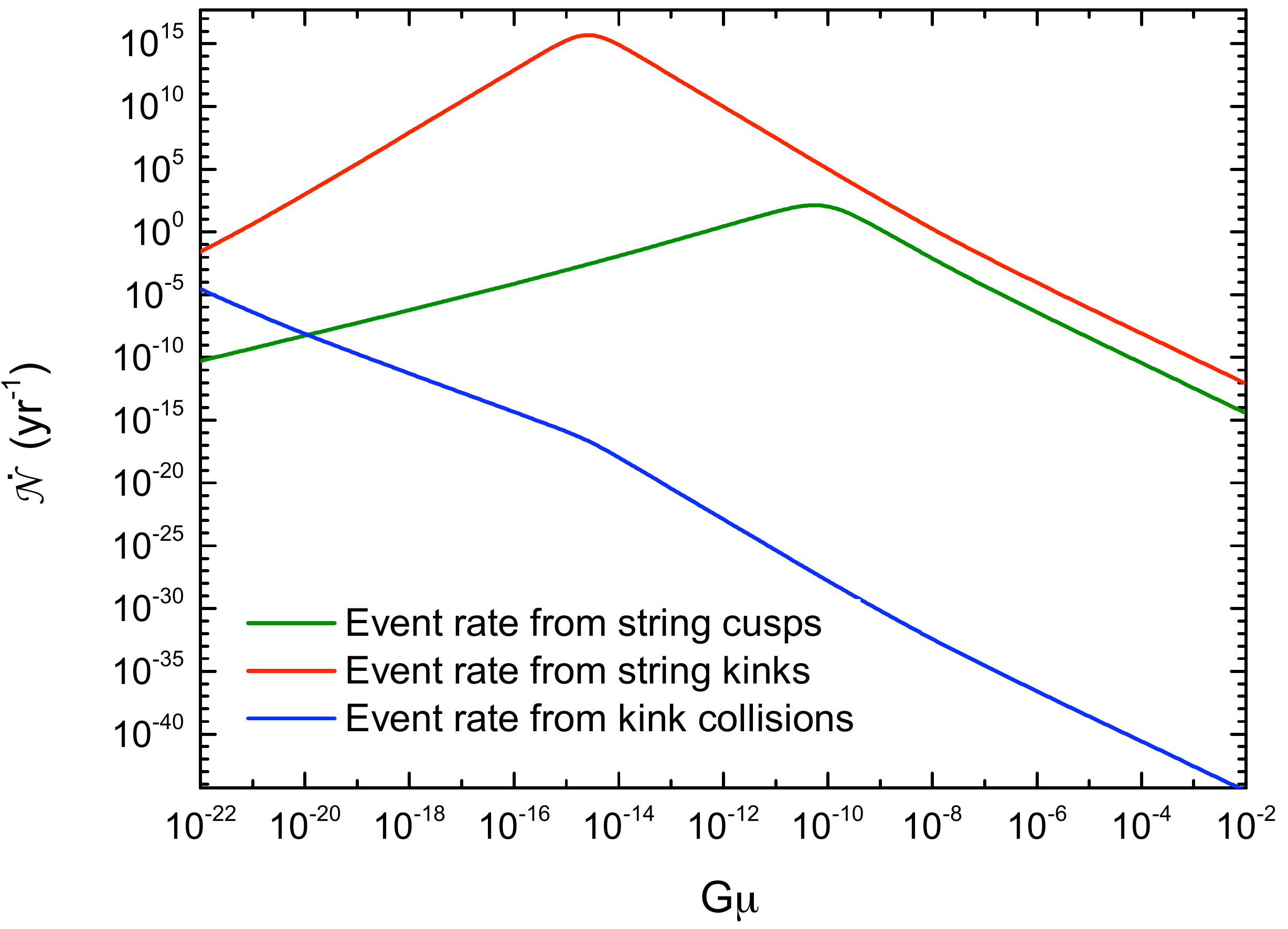}
\caption{The event rate of radio bursts from superconducting string 
loops with fixed observed frequency, $\nu_o = 1.23 {\rm GHz}$ as 
functions of $G \mu$. The current is taken to be 
$I=10^5~{\rm GeV}$. The green curve (middle curve at high 
$G \mu$) corresponds to the case of cusps; the red curve 
with largest event rate corresponds to the case of kinks; 
the blue curve corresponds to the case of kink-kink collision 
and gives a sub-dominant contribution at large $G \mu$.
The event rates for kinks and kink-kink collisions
should be rescaled by $N$ and $N^2$ if there are $N$ kinks
per loop. 
}
\label{Fig:Event_Gamma}
\end{figure}
Fig. \ref{Fig:Event_I} shows the total event rate dependence on the current,
$\curr$, as a power law. From Fig. \ref{Fig:Event_I}, we find the asymptotic
power law fits at large current values for the event rate of radio bursts from
cusps and kinks as follows,
\begin{numcases}
{\dot{\mathcal{N}} \simeq}
4.8 \times 10^{2}\, \curr_{5}^{2/3}\, {\rm yr}^{-1} \ ,
& {\rm (cusps)}\,  
\\
1.1\times 10^{5}\, N\, \curr_{5}^{1/2}\,  {\rm yr}^{-1} \ , 
& {\rm (kinks)}\,  
\end{numcases}
where $\curr_{5} \equiv \curr/({\rm 10^{5}\, GeV})$. In the figure, we did not
show the dependence of the current in the case of kink-kink collisions since
this case do not produce observable signal. 

Note that the plots in Figs.~1-3, are obtained by integrating over
the duration, which is theoretically constrained to be in the interval 
$\Delta \in (\Delta_{\min},\Delta_{\max})$
[see the discussion below Eq.~(\ref{Delta-rad})], and also experimentally
constrained due to the sensitivity of a particular radio transient search.
For example, for the Parkes survey, $\Delta_{\min} = 10^{-3}$. In 
the numerical evaluation of the event rate we have used the intersection 
of the theoretical and experimental ranges of the duration. 

%
Fig.~\ref{Fig:Event_Gamma} shows the total event rate dependence on 
$G\mu$. The event rate in the case of cusps and kinks
has a maximum at particular values of $G\mu$, namely, 
$G \mu \sim 10^{-10}$ for cusps and $G\mu \sim 10^{-15}$ for kinks. 
The maximum occurs because at small string tension electromagnetic 
radiation is the dominant energy loss mechanism, while at large 
tension gravitational losses dominate. Further, the event
rate from cusps and kinks depend differently on the duration
and flux --- see Eq.~(\ref{horror}) --- which causes their curves in 
the integrated event rate in Fig.~\ref{Fig:Event_Gamma} to bend over 
at different values of the tension, depending on the regimes $L(S,\Delta) \gg \Gamma G \mu t $ or $L(S, \Delta) \ll \Gamma G \mu t$.


\section{Conclusions}
\label{conclusions}

Current carrying superconducting cosmic strings will give three kinds
of transient electromagnetic bursts.
Bursts from cusps are strong and highly beamed, while bursts from kinks 
are weaker and less beamed, and the bursts from kink-kink collisions 
are weakest and not beamed. Only the bursts from cusps and kinks 
are strong enough to be observed. 

The bursts from cusps and kinks occur in all frequency bands but 
the width of the beams falls off with frequency. Thus the beams are 
wide in radio, and thin in gamma rays. So the event rate for bursts 
is largest in the radio bands, which is why the search for radio 
transients is the most likely to find bursts from superconducting 
strings. 

The search for radio bursts involves several parameters. First 
is the frequency at which observations are carried out, second 
is the lower cutoff in flux that the experiment is sensitive to, 
and the third is the duration of the burst. If we assume some
canonical values for the radio frequency and range of burst
durations, we can predict the event rate of radio transients from
superconducting strings as seen in Fig.~\ref{Fig:Event_S}. The
event rate is quite high, at the level of several a day at
1 Jy flux for the choice of string parameters,
and should be within easy reach of current efforts. If no bursts
are seen, their absence can be used to constrain the string parameters $G \mu$ and $\curr$.

Bursts from superconducting strings can be distinguished from
bursts from astrophysical sources as they are linearly polarized,
and should have characteristic frequency dependence. In principle,
the radio burst is accompanied by bursts at other frequencies 
but, since the beams at higher frequencies are narrower, they 
can miss detection. A radio burst should also be accompanied by
 gravitational wave bursts, but this can be very hard to
detect if the string is light. In some string models, there 
should also be an accompanying burst of neutrinos.

In our analysis, we have made some assumptions that we now 
spell out. The first is that all strings carry the same uniform
current. This assumes that the cosmological medium is magnetized
and that the current on strings has built up to its microscopically
determined saturation value. If there is a distribution of currents,
there will be an additional variable in the distribution of bursts
that can affect the event rate. A second assumption is that the
radiation backreaction does not drastically change the network 
properties. We have already accounted for some effects of 
backreaction. For example, loops evaporate with a certain lifetime
due to radiation. However, it is also possible that backreaction
prevents cusps from reappearing in every loop oscillation, or
that kinks smooth out very rapidly. We have not considered
these effects.


\acknowledgments
CYF thanks Xinmin Zhang and the Theory Division of the Institute for High Energy Physics, ES thanks APC, Paris, and DAS is grateful to ASU and CERN for hospitality while this work was being completed. This work was supported by the Department of Energy and by the National Science Foundation grant No. PHY-0854827 at ASU, and a CNRS PEPS grant between APC and ASU. 

\bibstyle{aps}

\end{document}